\documentclass[aps,onecolumn,letterpaper,oneside,preprint,tightenlines,draft,%
nobibnotes,nofootinbib,amsfonts,amssymb,amsmath,eqsecnum,showpacs,%
showkeys]{revtex4}

\newcommand{\cN}{\mathcal{N}}
\newcommand{\cV}{\mathcal{V}}
\newcommand{\cW}{\mathcal{W}}
\newcommand{\bxi}{\boldsymbol{\xi}}
\newcommand{\bzeta}{\boldsymbol{\zeta}}
\newcommand{\bvphi}{\boldsymbol{\varphi}}
\newcommand{\bW}{\boldsymbol{W}}
\newcommand{\bLamb}{\boldsymbol{\Lambda}}
\newcommand{\bOm}{\boldsymbol{\Omega}}
\newcommand{\tH}{\tilde{H}}
 
\newcommand{\tO}{\tilde{O}}
\newcommand{\tP}{\tilde{P}}

\newcommand{\tcV}{\tilde{\cV}}

\newcommand{\tvph}{\tilde{\varphi}}

\newcommand{\bbN}{\mathbb{N}}

\newcommand{\bbC}{\mathbb{C}}
\newcommand{\rme}{\mathrm{e}}
\newcommand{\rmd}{\mathrm{d}}
\newcommand{\rmA}{\mathrm{A}}
\newcommand{\rmT}{\mathrm{T}}
\newcommand{\braket}[1]{\bigl\langle{#1}\bigr\rangle}
\newcommand{\fsl}{\mathfrak{sl}} 
\newcommand{\htcV}{\widehat{\tcV}}
\newcommand{\htO}{\widehat{\tO}}
\newcommand{\htH}{\widehat{\tH}}

\newcommand{\htP}{\widehat{\tP}}
\newcommand{\htvph}{\widehat{\tvph}}
\newcommand{\hbP}{\widehat{\bar{P}}}
\newcommand{\hbcV}{\widehat{\bar{\cV}}}
\newcommand{\hA}{\widehat{A}}
\newcommand{\hB}{\widehat{B}}
\newcommand{\hC}{\widehat{C}}
\newcommand{\hE}{\widehat{E}}
\newcommand{\hF}{\widehat{F}}
\newcommand{\hQ}{\widehat{Q}}
\newcommand{\hW}{\widehat{W}}
\newcommand{\blambda}{\bar{\lambda}}

\newcommand{\hbOm}{\widehat{\bOm}}
\newcommand{\Tr}{\operatorname{Tr}}

\begin{document}


%
%

\title{$GL(3,\bbC)$ Invariance of Type B 3-fold Supersymmetric Systems}
\author{Toshiaki Tanaka}
\email{tanaka.toshiaki@ocha.ac.jp}
\affiliation{Department of Physics,
 Faculty of Science, Ochanomizu University,
 2-1-1 Ohtsuka, Bunkyo-ku, Tokyo 112-8610, Japan}


\begin{abstract}

Type B 3-fold supersymmetry is a necessary and sufficient condition for
a quantum Hamiltonian to admit three linearly independent local solutions
in closed form. We show that any such a system is invariant under
$GL(3,\bbC)$ homogeneous linear transformations. In particular, we prove
explicitly that the parameter space is transformed as an adjoint
representation of it and that every coefficient of the characteristic
polynomial appeared in 3-fold superalgebra is algebraic invariants.
In the type A case, it includes as a subgroup the $GL(2,\bbC)$ linear
fractional transformation studied in the literature.
We argue that any $\cN$-fold supersymmetric system has a $GL(\cN,\bbC)$
invariance for an arbitrary integral $\cN$.

\end{abstract}


\pacs{02.10.Ud; 02.20.Qs; 03.65.Ca; 11.30.Pb}
\keywords{$\cN$-fold supersymmetry; Schr\"{o}dinger operators; Linear
transformations; Group representations; Algebraic invariants;
Quasi-solvability}




\maketitle

\section{Introduction}
\label{sec:intro}

One of the most important roles of mathematical science is to unveil
universal mathematical structure which is hidden in various models.
For this aim, the concept of symmetry has played a central and crucial
role, see, for instance, a classical reference~\cite{Co85}. In many
cases, one would first consider a symmetry, regardless of it is exact
or only approximate, which characterize a system under consideration,
and then build a corresponding model such that the symmetry is realized
manifestly or may be broken dynamically in it. In other cases, however,
a symmetry is discovered a posteriori or even accidentally in an
existing model. Recently developed \emph{$\cN$-fold supersymmetry
(SUSY)} is such an example, which has been recognized that it exists
in any (quasi-)solvable quantum one-body Hamiltonian since its
equivalence to weak quasi-solvability was proved in~\cite{AST01b}.
It means in particular that any second-order linear ordinary
differential equation has this symmetry provided that it admits
a certain number of local solutions in closed form, since such
an operator can be always transformed into a Schr\"{o}dinger one.
For a review of $\cN$-fold SUSY, see Ref.~\cite{Ta09}.

In our previous paper~\cite{Ta13}, we showed that a particular type
of $\cN$-fold SUSY, called \emph{type B}, for $\cN_3$
is a necessary and sufficient condition for a quantum Hamiltonian to
admit three linearly independent local solutions in closed form, and
obtained its most general analytical form. Contrary to the virtually
every exactly solvable model whose analytical solutions are expressible
in terms of a polynomial system in a certain variable, type B systems
are peculiar in the sense that their solutions are essentially of
non-polynomial character. In addition, they are not Lie-algebraic,
in contrast with many quasi-solvable systems like the famous $\fsl(2)$
ones constructed in~\cite{Tu88} and others~\cite{KO90,GKO94b}.
Hence, most of their aspects have still remained unknown and uncovered.

One of issues to be interested in next is, what kind of potentials is
realizable in type B 3-fold SUSY. In the case of type A $\cN$-fold
SUSY, the problem was entirely solved by considering the $GL(2,\bbC)$
symmetry in~\cite{Ta03a}. The use of the latter transformation group
is originated from the investigation into the normalizability of
$\fsl(2)$ Lie-algebraic quasi-solvable models in~\cite{GKO93}.
The essential point is that one can classify the models by considering
the inequivalent classes with respect to the symmetry transformation.
Later in~\cite{Ta03a}, it was recognized that the most general type A
systems were invariant under the same $GL(2,\bbC)$ transformation and
thus that they were classified in the same way as done in~\cite{GKO93}.
Hence, it is reasonable to expect that discovery of symmetry in the
most general type B 3-fold SUSY models would provide one a useful and
powerful tool to examine the classification problem.

In this paper, we explore in detail effect of a general variable
transformation on a type B 3-fold SUSY system and its invariance under
a $GL(3,\bbC)$ homogeneous linear transformation. For the purpose, we
first introduce a change of variable in a three-dimensional linear
function space in a gauged space which entirely characterize the system
under consideration. We then examine the resultant transformations of
all the constituent functions and operators involved in the system.
Based on these arrangements, we proceed to a study of a $GL(3,\bbC)$
transformation which preserves the linear space. Due to the latter
property, every component quantities of the system has invariance
under the transformation. After completing the full analysis, we
consider a type A 3-fold SUSY system as a particular case of the type B.
In particular, we reconsider the $GL(2,\bbC)$ invariance of any type A
model in~\cite{Ta03a} and show that it is a subgroup of the $GL(3,\bbC)$
in this work. Hence, we reproduce all the results derived in the latter
reference. Finally, we generalize the argument to assert that any
$\cN$-fold SUSY system has a $GL(\cN,\bbC)$ invariance.

We organize the paper as follows. In Section~\ref{sec:InB3}, we briefly
summarize the ingredients of type B 3-fold SUSY which constitute
a basis for the present purpose. In Section~\ref{sec:GTB3}, we develop
effects of a general variable transformation in type B 3-fold SUSY.
We present transformation formulas for various quantities there.
In Section~\ref{sec:GL3C}, we proceed to a homogeneous linear
transformation $GL(3,\bbC)$ as a particular case of the general
one in Section~\ref{sec:GTB3}. We show explicitly that any type B
3-fold SUSY system is invariant under the $GL(3,\bbC)$ transformation
and that there exist three invariants of the transformation, in terms
of which both the Hamiltonians and 3-fold supercharge components are
entirely expressible. We prove that the parameter space, which is dual
to the linear space of quasi-solvable operators, is transformed as an
adjoint representation of $GL(3,\bbC)$. In Section~\ref{sec:B3SA},
we present a product of the type B 3-fold supercharges which constitutes
a part of type B 3-fold superalgebra and provides a characteristic
polynomial for the spectrum. We show that its coefficients are all
invariants of the $GL(3,\bbC)$ transformation. In Section~\ref{sec:Alim},
we consider type A 3-fold SUSY as a special case of the most general
type B. In particular, we show that the $GL(2,\bbC)$ linear fractional
transformation in~\cite{Ta03a} is a subgroup of the $GL(3,\bbC)$.
In Section~\ref{sec:GLNC}, we argue by following the same reasoning
as in Section~\ref{sec:GL3C} that any $\cN$-fold SUSY system has
a $GL(\cN,\bbC)$ invariance for an arbitrary $\cN\in\bbN$. In
Section~\ref{sec:discus}, we summarize the results and discuss
prospects in the future development.

\section{Ingredients of Type B 3-fold SUSY}
\label{sec:InB3}

Type B $\cN$-fold SUSY was first discovered in Ref.~\cite{GT04} by a simple
deformation of type A $\cN$-fold supercharge.
The component of type B 3-fold supercharge is given by
\begin{align}
P_{3}^{-}=\left(\frac{\rmd}{\rmd q}+W(q)-E(q)-F(q)\right)\left(
 \frac{\rmd}{\rmd q}+W(q)\right)\left(\frac{\rmd}{\rmd q}+W(q)+E(q)\right),
\label{eq:P3-}
\end{align}
where the three functions $W(q)$, $E(q)$, and $F(q)$ are at present
arbitrary. A pair of Hamiltonians
\begin{align}
H^{\pm}=-\frac{1}{2}\frac{\rmd^{2}}{\rmd q^{2}}+V^{\pm}(q),
\label{eq:H+-}
\end{align}
is said to have type B $3$-fold SUSY if it is intertwined by $P_{3}^{-}$
in (\ref{eq:P3-}) as
\begin{align}
P_{3}^{\mp}H^{\mp}=H^{\pm}P_{3}^{\mp},
\label{eq:inter}
\end{align}
where $P_{3}^{+}$ is the transposition of $P_{3}^{-}$ in the $q$-space, 
$P_{3}^{+}=(P_{3}^{-})^{\rmT}$. It was shown that the intertwining
relation (\ref{eq:inter}) holds if and only if the potential
terms in (\ref{eq:H+-}) have the following form
\begin{align}
V^{\pm}=\frac{1}{2}W^{2}-\frac{1}{3}\left( 2 E'-E^{2}\right)-\frac{1}{6}
 \left( 2 F'+2 W\! F-2 E F-F^{2}\right)\pm\frac{1}{2}\left(3 W'-F'\right),
\label{eq:cond1}
\end{align}
and simultaneously the three functions $W(q)$, $E(q)$, and $F(q)$ satisfy
\begin{align}
\left(\frac{\rmd}{\rmd q}-E\right) F'_{1}-\frac{F}{2}\left( F'_{1}
 -\frac{F'_{2}}{6}\right) =0,
\label{eq:cond2}\\
\left(\frac{\rmd}{\rmd q}-2 E-\frac{3}{2}F\right)\left(\frac{\rmd}{\rmd q}
 -E\right) F'_{2}+\frac{3}{2}\left( 2 F'-2 E F-F^{2}\right)\left( F'_{1}
 -\frac{F'_{2}}{6}\right) =0,
\label{eq:cond3}
\end{align}
where $F_{1}(q)$ and $F_{2}(q)$ are given by
\begin{align}
\begin{split}
F_{1}&=W'+E W-\frac{1}{4}\left( F'-2 W\! F+2 E F+F^{2}\right),\\
F_{2}&=E'+E^{2}+\frac{1}{2}\left( F'-2 W\! F+2 E F+F^{2}\right).
\end{split}
\label{eq:F1F2}
\end{align}
They can be easily integrated analytically, but the problem gets simpler
if we make a change of variable $z=z(q)$ and introduce three functions
$f(z)$, $A(z)$, and $Q(z)$ defined by
\begin{align}
\begin{split}
&2 A(z(q))=z'(q)^{2},\qquad E(q)=\frac{z''(q)}{z'(q)},\\
&F(q)=\frac{f'''(z(q))}{f''(z(q))}z'(q),\qquad W(q)=-\frac{Q(z(q))}{z'(q)}.
\label{eq:fAQ}
\end{split}
\end{align}
Then, the pair of type B $3$-fold SUSY Hamiltonians $H^{\pm}$ having
the potential terms (\ref{eq:cond1}) can be written as
\begin{align}
H^{\pm}=\rme^{-\cW_{3}^{\pm}(q)}\bar{\tH}[z]\,\rme^{\cW_{3}^{\pm}(q)}
 \Bigr|_{z=z(q)},
\label{eq:btH}
\end{align}
where the gauged Hamiltonian $\tH^{-}$ is a second-order linear differential
operator
\begin{align}
\tH^{-}[z]=-A(z)\frac{\rmd^{2}}{\rmd z^{2}}-B(z)\frac{\rmd}{\rmd z}-C(z),
 \qquad B(z)=Q(z)-\frac{A'(z)}{2},
\label{eq:tH-B}
\end{align}
and the coefficients $A(z)$, $B(z)$, and $C(z)$ are functions of $z$
given by
\begin{align}
A(z)f''(z)=&\;\left[(c_{2}z-b_{2})f(z)+c_{1}z^{2}+(c_{0}-b_{1})z-b_{0}
 \right]f'(z)\notag\\
&\;-[c_{2}f(z)+c_{1}z+c_{0}-a_{2}]f(z)+a_{1}z+a_{0},
\label{eq:Az}\\
B(z)=&\;-(c_{2}z-b_{2})f(z)-c_{1}z^{2}-(c_{0}-b_{1})z+b_{0},\\
C(z)=&\;c_{2}f(z)+c_{1}z+c_{0}.
\label{eq:Cz}
\end{align}
In the gauged $z$-space, the component of type B $3$-fold supercharge
reads as
\begin{align}
\tP_{3}^{-}[z]=\rme^{\cW_{3}^{-}}P_{3}^{-}\rme^{-\cW_{3}^{-}}
 =z'(q)^{3}\left(\frac{\rmd}{\rmd z}-\frac{f'''(z)}{f''(z)}\right)
 \frac{\rmd^{2}}{\rmd z^{2}},
\label{eq:tP3-}
\end{align}
and it annihilates all the elements of a three-dimensional linear
function space $\tcV_{3}^{-}$:
\begin{align}
\tcV_{3}^{-}[z]=\ker \tP_{3}^{-}[z]=\braket{1,z,f(z)}.
\label{eq:tV3-}
\end{align}
On the other hand, it follows from the gauge-transformed version of
the intertwining relation (\ref{eq:inter}) with upper signs, namely, 
$\tP_{3}^{-}\tH^{-}=\tH^{+}\tP_{3}^{-}$, that $\tH^{-}\ker \tP_{3}^{-}
\subset\ker \tP_{3}^{-}$. Hence, any gauged type B $3$-fold SUSY
Hamiltonian $\tH^{-}$ in (\ref{eq:tH-B}) with (\ref{eq:Az})--(\ref{eq:Cz})
preserves the linear space (\ref{eq:tV3-}), which is thus called
a \emph{solvable sector} of $\tH^{-}$.

A solvable sector $\bar{\cV}_{3}^{+}$ preserved by $\bar{H}^{+}$ is 
characterized by the gauge-transformed operator of the transposed
component $P_{3}^{+}$ of type B $3$-fold supercharge
\begin{align}
\bar{P}_{3}^{+}[z]=\rme^{\cW_{3}^{+}}P_{3}^{+}\rme^{-\cW_{3}^{+}}
 =-z'(q)^{3}\frac{\rmd^{2}}{\rmd z^{2}}\left(\frac{\rmd}{\rmd z}+\frac{
 f'''(z)}{f''(z)}\right).
\label{eq:bP3+}
\end{align}
In fact, the gauge transformation with $\cW_{3}^{+}$ of the intertwining
relation (\ref{eq:inter}) with lower signs, namely, 
$\bar{P}_{3}^{+}\bar{H}^{+}=\bar{H}^{-}\bar{P}_{3}^{+}$, immediately
leads to $\bar{H}^{+}\ker \bar{P}_{3}^{+}\subset\ker\bar{P}_{3}^{+}$.
Hence, we obtain from (\ref{eq:bP3+})
\begin{align}
\bar{\cV}_{3}^{+}[z]=\ker \bar{P}_{3}^{+}[z]=\frac{1}{f''(z)}
 \braket{1,f'(z),z f'(z)-f(z)}.
\label{eq:bV3+}
\end{align}

\section{General Transformation in Type B $3$-fold SUSY}
\label{sec:GTB3}

Let us first review and then develop further the effect of a general
transformation on the three-dimensional type B space studied in
Ref.~\cite{Ta13}, Section~5:
\begin{align}
\tcV_{3}^{-}[z]=\braket{1,z,f(z)}\ \to\ \htcV{}_{3}^{-}[w]=
 \braket{\tvph_{1}(w),\tvph_{2}(w),\tvph_{3}(w)}.
\label{eq:trB}
\end{align}
Since the latter space is expressed as
\begin{align}
\htcV{}_{3}^{-}[w]=\tvph_{1}(w)\tcV_{3}^{-}[z]\Bigr|_{z=\tvph_{2}/
 \tvph_{1},\,f(z)=\tvph_{3}/\tvph_{1}},
\end{align}
an operator of a certain property acting on it is transformed as
\begin{align}
\tO[z]\ \to\ \htO[w]=\tvph_{1}(w)\tO[z]\tvph_{1}(w)^{-1}
 \Bigr|_{z=\tvph_{2}/\tvph_{1},\,f(z)=\tvph_{3}/\tvph_{1}}.
\label{eq:opt}
\end{align}
Several useful formulas needed for performing a transformation are
summarized in Appendix~\ref{app:forms}. For instance,
the gauged type B $3$-fold supercharge component $\tP_{3}^{-}$ is
transformed as
\begin{align}
\htP{}_{3}^{-}[w]=&\;w'(q)^{3}\left( \frac{\rmd}{\rmd w}
 +\frac{\tvph'_{1}(w)}{\tvph_{1}(w)}+\frac{W'_{2,1}(w)}{W_{2,1}(w)}
 -\frac{W'_{31,21}(w)}{W_{31,21}(w)}\right)\notag\\
&\times\left( \frac{\rmd}{\rmd w}+\frac{\tvph'_{1}(w)}{\tvph_{1}(w)}
 -\frac{W'_{2,1}(w)}{W_{2,1}(w)}\right)
 \left( \frac{\rmd}{\rmd w}-\frac{\tvph'_{1}(w)}{\tvph_{1}(w)}\right),
\label{eq:tP3-'}
\end{align}
where the Wronskians $W_{i,j}(w)$ and $W_{ij,kl}(w)$ are defined in
(\ref{eq:Wron1}) and (\ref{eq:Wron3}).
The plus component of the gauged type B $3$-fold supercharge
$\hbP{}_{3}^{+}[w]$ and the linear space $\hbcV{}_{3}^{+}[w]$
annihilated by it in the $w$-space are related to the corresponding
quantities in the $z$-space, $\bar{P}_{3}^{+}[z]$ in (\ref{eq:bP3+})
and $\bar{\cV}_{3}^{+}[z]$ in (\ref{eq:bV3+}), respectively, as
\begin{align}
\begin{split}
\hbP{}_{3}^{+}[w]&=\tvph_{1}(w)^{3}W_{2,1}(w)^{-2}\bar{P}_{3}^{+}[z]
 W_{2,1}(w)^{2}\tvph_{1}(w)^{-3}\bigl|_{z=\tvph_{2}/\tvph_{1}, f(z)=
 \tvph_{3}/\tvph_{1}}\,,\\
\hbcV{}_{3}^{+}[w]&=\tvph_{1}(w)^{3}W_{2,1}(w)^{-2}\bar{\cV}_{3}^{+}[z]
 \bigl|_{z=\tvph_{2}/\tvph_{1}, f(z)=\tvph_{3}/\tvph_{1}}\,.
\end{split}
\end{align}
With these formulas, we obtain
\begin{align}
\hbP{}_{3}^{+}[w]=&-w'(q)^{3}\left(\frac{\rmd}{\rmd w}
 +\frac{\tvph'_{1}(w)}{\tvph_{1}(w)}\right)\left(
 \frac{\rmd}{\rmd w}-\frac{\tvph'_{1}(w)}{\tvph_{1}(w)}
 +\frac{W'_{2,1}(w)}{W_{2,1}(w)}\right)\notag\\
&\times\left(\frac{\rmd}{\rmd w}-\frac{\tvph'_{1}(w)}{\tvph_{1}(w)}
 -\frac{W'_{2,1}(w)}{W_{2,1}(w)}+\frac{W'_{31,21}(w)}{W_{31,21}(w)}\right),
\label{eq:bP3+'}
\end{align}
and
\begin{align}
\hbcV{}_{3}^{+}[w]=\frac{\tvph_{1}(w)}{W_{31,21}(w)}
 \braket{W_{2,1}(w), W_{3,1}(w), W_{3,2}(w)}.
\label{eq:bV3+'}
\end{align}

We shall next consider the effect of the above transformation in the
original $q$-space. For this purpose, we must know the transformations
of the functions $E(q)$, $F(q)$, and $W(q)$ introduced in (\ref{eq:fAQ}).
In terms of the new variable $w(q)$, the corresponding transformed
functions $\hE(q)$, $\hF(q)$, and $\hW(q)$ are expressed as
\begin{align}
\hE(q)=\frac{w''(q)}{w'(q)},\quad \hF(q)=\frac{f'''(w(q))}{f''(w(q))}
 w'(q),\quad \hW(q)=-\frac{\hQ(w(q))}{w'(q)}.
\end{align}
To begin with, we note that with the aid of the formulas (\ref{eq:f2}),
we have
\begin{align}
\begin{split}
E(q)&=\frac{w''(q)}{w'(q)}+\left(\frac{W'_{2,1}(w(q))}{W_{2,1}(w(q))}
 -\frac{2\tvph'_{1}(w(q))}{\tvph_{1}(w(q))}\right)w'(q),\\
F(q)&=\left(\frac{W'_{31,21}(w(q))}{W_{31,21}(w(q))}
 -\frac{3W'_{2,1}(w(q))}{W_{2,1}(w(q))}
 +\frac{2\tvph'_{1}(w(q))}{\tvph_{1}(w(q))}\right)w'(q).
\end{split}
\label{eq:t0}
\end{align}
Hence, the transformation of $E(q)$ immediately reads as
\begin{align}
\hE(q)=E(q)-\left(\frac{W'_{2,1}(w(q))}{W_{2,1}(w(q))}
 -\frac{2\tvph'_{1}(w(q))}{\tvph_{1}(w(q))}\right)w'(q).
\label{eq:t1}
\end{align}
On the other hand, the transformation of $F(q)$ cannot be determined
unless the dependence of $\tvph_{i}(w)$ ($i=1,2,3$) on the function
$f(w)$ is specified, which we shall later consider.

Next, to establish the transformation of the function $W(q)$, we must
first know the transformation of $Q(z)$ introduced in (\ref{eq:tH-B}).
The transformed functions $\hA(w)$ and $\hQ(w)$ are thus defined as
\begin{align}
\htH{}^{-}[w]=-\hA(w)\frac{\rmd^{2}}{\rmd w^{2}}-\left(\hQ(w)
 -\frac{\hA{}'(w)}{2}\right)\frac{\rmd}{\rmd w}-\hC(w).
\label{eq:htH-1}
\end{align}
Applying the formula (\ref{eq:opt}) to $\tH^{-}[z]$, we have
\begin{align}
\htH{}^{-}[w]=&-A(z)\left(\frac{\rmd w}{\rmd z}\right)^{2}\frac{
 \rmd^{2}}{\rmd w^{2}}+\biggl[2A(z)\left(\frac{\rmd w}{\rmd z}\right)^{2}
 \frac{\tvph'_{1}}{\tvph_{1}}-A(z)\frac{\rmd^{2}w}{\rmd z^{2}}\notag\\
&-\left(Q(z)-\frac{A'(z)}{2}\right)\frac{\rmd w}{\rmd z}\biggr]
 \frac{\rmd}{\rmd w}+A(z)\left(\frac{\rmd w}{\rmd z}\right)^{2}
 \left(\frac{\tvph''_{1}}{\tvph_{1}}
 -2\frac{(\tvph'_{1})^{2}}{(\tvph_{1})^{2}}\right)\notag\\
&+\left[A(z)\frac{\rmd^{2} w}{\rmd z^{2}}+\left(Q(z)-\frac{A'(z)}{2}
 \right)\frac{\rmd w}{\rmd z}\right]\frac{\tvph'_{1}}{\tvph_{1}}-C(z)
 \Biggr|_{z=\tvph_{2}/\tvph_{1},f(z)=\tvph_{3}/\tvph_{1}}.
\label{eq:htH-2}
\end{align}
Comparing the (\ref{eq:htH-1}) and (\ref{eq:htH-2}), and using
(\ref{eq:f1}), we obtain the transformation rule as
\begin{align}
\begin{split}
&\hA(w)=A(z)\frac{(\tvph_{1})^{4}}{(W_{2,1})^{2}},\qquad
 \hB(w)=B(z)\frac{(\tvph_{1})^{2}}{W_{2,1}}-A(z)
 \frac{W'_{2,1}(\tvph_{1})^{4}}{(W_{2,1})^{3}},\\
&\hC(w)=C(z)-B(z)\frac{\tvph_{1}\tvph'_{1}}{W_{2,1}}
 +A(z)\frac{W_{2',1'}(\tvph_{1})^{4}}{(W_{2,1})^{3}},\\
&\hQ(w)=Q(z)\frac{(\tvph_{1})^{2}}{W_{2,1}}-2A(z)\left(\frac{W'_{2,1}}{
 W_{2,1}}-\frac{\tvph'_{1}}{\tvph_{1}}\right)\frac{(\tvph_{1})^{4}}{
 (W_{2,1})^{2}}.
\end{split}
\label{eq:tABCQ}
\end{align}
The last formula determines the transformation of $W(q)$ as
\begin{align}
\hW(q)=W(q)+\left(\frac{W'_{2,1}(w(q))}{W_{2,1}(w(q))}
 -\frac{\tvph'_{1}(w(q))}{\tvph_{1}(w(q))}\right)w'(q).
\label{eq:t2}
\end{align}

For a concrete calculation of the transformation in (\ref{eq:tABCQ}),
it is convenient to rewrite the functions $A(z)$, $B(z)$, and $C(z)$
given in (\ref{eq:Az})--(\ref{eq:Cz}) in matrix form. Let us arrange
the set of parameters $\{a_{i},b_{i},c_{i}\}$ in a three-by-three
matrix $\bOm$ and the bases of $\tcV_{3}^{-}[z]$ and $\htcV{}_{3}^{-}[w]$
in three-component column vectors $\bvphi_{0}(z)$ and $\bvphi(w)$,
respectively, as
\begin{align}
\bOm=\left(
 \begin{array}{ccc}
 c_{0} & c_{1} & c_{2}\\
 b_{0} & b_{1} & b_{2}\\
 a_{0} & a_{1} & a_{2}
 \end{array}\right),\quad
\bvphi_{0}(z)=\left(\begin{array}{c}1\\ z\\ f(z)\end{array}\right),\quad
\bvphi(w)=\left(\begin{array}{c}\tvph_{1}(w)\\ \tvph_{2}(w)\\ \tvph_{3}(w)
\end{array}\right).
\end{align}
Then, the formulas (\ref{eq:Az})--(\ref{eq:Cz}) are rewritten as
\begin{align}
\begin{split}
A(z)f''(z)&=\bxi^{\rmT}(z)\bOm\bvphi_{0}(z),\\
B(z)&=-\bzeta_{0}^{\rmT}(z)\bOm\bvphi_{0}(z),\\
C(z)&=\bzeta_{0}^{\prime\rmT}(z)\bOm\bvphi_{0}(z),
\end{split}
\label{eq:ABC1}
\end{align}
where the superscript $\rmT$ denotes the transposition of a matrix, and
the column vectors $\bxi(z)$ and $\bzeta_{0}(z)$ are defined by
\begin{align}
\bxi(z)=\left(\begin{array}{c}zf'(z)-f(z)\\ -f'(z)\\ 1\end{array}\right),
 \qquad\bzeta_{0}(z)=\left(\begin{array}{c}z\\ -1\\ 0\end{array}\right).
\end{align}
Applying the formulas
(\ref{eq:f0}) and (\ref{eq:f1}), they are expressed in terms of the new
variable $w$ as
\begin{align}
\begin{split}
A(z)f''(z)&=\frac{1}{W_{2,1}(w)\tvph_{1}(w)}\bW^{\rmT}(w)\bOm\bvphi(w),\\
B(z)&=-\frac{1}{\tvph_{1}(w)^{2}}\bzeta^{\rmT}(w)\bOm\bvphi(w),\\
C(z)&=\frac{1}{\tvph_{1}(w)}\bzeta_{0}^{\prime\rmT}(w)\bOm\bvphi(w),
\end{split}
\end{align}
where
\begin{align}
 \bW(w)=\left(\begin{array}{c}W_{3,2}(w)\\ -W_{3,1}(w)\\ W_{2,1}(w)
 \end{array}\right),\qquad
 \bzeta(w)=\left(\begin{array}{c}\tvph_{2}(w)\\ -\tvph_{1}(w)\\ 0
 \end{array}\right).
\end{align}
Substituting them into (\ref{eq:tABCQ}), we obtain the transformed functions 
$\hA(w)$, $\hB(w)$, and $\hC(w)$ in matrix form as
\begin{align}
\begin{split}
\hA(w)f''(w)&=\frac{\tvph_{1}(w)}{W_{31,21}(w)}\bW^{\rmT}(w)\bOm\bvphi(w),\\
\hB(w)&=-\frac{\tvph_{1}(w)}{W_{31,21}(w)}\bW^{\prime\rmT}(w)\bOm\bvphi(w),\\
\hC(w)&=\frac{\tvph_{1}(w)}{W_{31,21}(w)}\bW_{\!\!\prime\prime}^{\rmT}(w)
 \bOm\bvphi(w),
\end{split}
\label{eq:hABC1}
\end{align}
where the column vector $\bW_{\!\!\prime\prime}(w)$ is defined by
\begin{align}
\bW_{\!\!\prime\prime}(w)=\left(\begin{array}{c}
 W_{3',2'}(w)\\ -W_{3',1'}(w)\\ W_{2',1'}(w)\end{array}\right).
\end{align}
The Wronskian $W_{i',j'}(w)$ appeared in the components is defined in
(\ref{eq:Wron2}).

\section{$GL(3,\bbC)$ Invariance of Type B $3$-fold SUSY}
\label{sec:GL3C}

We are now in a position to consider a homogeneous linear transformation
$GL(3,\bbC)$ of the basis:
\begin{align}
\bvphi(w)=\bLamb\bvphi_{0}(w),\qquad\bLamb=\left(
 \begin{array}{ccc}\lambda_{11} & \lambda_{12} & \lambda_{13}\\
 \lambda_{21} & \lambda_{22} & \lambda_{23}\\
 \lambda_{31} & \lambda_{32} & \lambda_{33}\end{array}\right)\in GL(3,\bbC),
\label{eq:GL3}
\end{align}
where $\lambda_{ij}\in\bbC$ ($i,j=1,2,3$) are all constants satisfying
$\det\bLamb\neq 0$.
It is evident that the three-dimensional type B space is invariant under
any of the $GL(3,\bbC)$ transformation, that is,
\begin{align}
\htcV{}_{3}^{-}[w]=\tcV_{3}^{-}[w].
\end{align}
It means in particular that type B 3-fold SUSY Hamiltonians $H^{\pm}$ and
supercharge component $P_{3}^{-}$ must have the same forms both in terms
of the variable $z(q)$ and in terms of $w(q)$. To see it, we firstly need
to know more explicit forms of the Wronskians in the $GL(3,\bbC)$ case.
We summarize them in (\ref{eq:f3}). We note that, applying them to
(\ref{eq:bV3+'}) and comparing the result with (\ref{eq:bV3+}), we can
also check the invariance of the other subspace:
\begin{align}
\hbcV{}_{3}^{+}[w]=\frac{1}{f''(w)}
 \braket{1, f'(w), w f'(w)-f(w)}=\bar{\cV}_{3}^{+}[w].
\end{align}
Substituting the formula for $W_{31,21}(w)$ into (\ref{eq:t0}), we
eventually obtain the transformation of the function $F(q)$:
\begin{align}
\hF(q)=F(q)+3\left(\frac{W'_{2,1}(w(q))}{W_{2,1}(w(q))}
 -\frac{\tvph'_{1}(w(q))}{\tvph_{1}(w(q))}\right)w'(q).
\label{eq:t3}
\end{align}
{}From the transformation formulas (\ref{eq:t1}), (\ref{eq:t2}), and
(\ref{eq:t3}) together with (\ref{eq:f3}), we can show that there are
three functions $I_{i}$ ($i=1,2,3$) composed of $E(q)$, $W(q)$, and $F(q)$ 
which are invariant under the $GL(3,\bbC)$ as the followings:
\begin{align}
\begin{split}
&I_{1}[E,W,F]=W-\frac{1}{3}F,\quad I_{2}[E,W,F]=2E'+F'-E^{2}-EF
 -\frac{1}{3}F^{2},\\
&I_{3}[E,W,F]=F''-E'F-3EF'-2FF'+2E^{2}F+2EF^{2}+\frac{4}{9}F^{3}.
\end{split}
\label{eq:inv}
\end{align}
That is, all the above satisfy
\begin{align}
I_{i}[\hE,\hW,\hF]=I_{i}[E,W,F],\qquad (i=1,2,3).
\end{align}
The invariance of $I_{1}$ is trivial from (\ref{eq:t2}) and (\ref{eq:t3}).
To show it for $I_{2}$ and $I_{3}$, we need the following identity which
is easily shown with the formula (\ref{eq:f3}):
\begin{align}
J(w)=(W''_{2,1}\tvph_{1}-W'_{2,1}\tvph'_{1}+W_{2,1}\tvph''_{1})f''
 -W'_{2,1}\tvph_{1}f'''=0.
\end{align}
Indeed, a direct calculation shows that
\begin{align*}
I_{2}[\hE,\hW,\hF]=&\;I_{2}[E,W,F]+\frac{J}{W_{2,1}\tvph_{1}f''}w'(q)^{2},\\
I_{3}[\hE,\hW,\hF]=&\;I_{3}[E,W,F]
 -6\frac{\tvph'''_{1}f''-\tvph''_{1}f'''}{\tvph_{1}f''}w'(q)^{3}\\
&+\frac{[3(W_{2,1}\tvph'_{1}-W'_{2,1}
 \tvph_{1})f''-5W_{2,1}\tvph_{1}f''']J+3W_{2,1}\tvph_{1}f''J'}{
 (W_{2,1})^{2}(\tvph_{1})^{2}(f'')^{2}}w'(q)^{3},
\end{align*}
and their invariance is now manifest since $\tvph'''_{1}f''=\tvph''_{1}f'''$
by (\ref{eq:f3}). Then, the type B 3-fold supercharge component $P_{3}^{-}$
in (\ref{eq:P3-}) and the pair of type B 3-fold SUSY potentials $V^{\pm}$
in (\ref{eq:cond1}) are all expressible solely in terms of these
invariants as
\begin{align}
P_{3}^{-}=&\;\frac{\rmd^{3}}{\rmd q^{3}}+3I_{1}(q)\frac{\rmd^{2}}{\rmd q^{2}}
 +\left[3I'_{1}(q)+3I_{1}(q)^{2}+I_{2}(q)\right]\frac{\rmd}{\rmd q}\notag\\
&+I''_{1}(q)+3I_{1}(q)I'_{1}(q)+I_{1}(q)^{3}+I_{1}(q)I_{2}(q)+\frac{1}{2}
 I'_{2}(q)-\frac{1}{6}I_{3}(q),
\label{eq:invP}\\
V^{\pm}=&\;\frac{1}{2}I_{1}(q)^{2}-\frac{1}{3}I_{2}(q)
 \pm\frac{3}{2}I'_{1}(q),
\label{eq:invV}
\end{align}
and hence their invariance under the $GL(3,\bbC)$ is clearly shown.
However, we note that each of the factors $P_{3i}^{-}$ ($i=1,2,3$) in
the type B 3-fold supercharge $P_{3}^{-}=P_{31}^{-}P_{32}^{-}P_{33}^{-}$,
where
\begin{align}
P_{31}^{-}=\frac{\rmd}{\rmd q}+W-E-F,\qquad P_{32}^{-}=\frac{\rmd}{\rmd q}
 +W,\qquad P_{33}^{-}=\frac{\rmd}{\rmd q}+W+E,
\end{align}
is not invariant; actually they transform as
\begin{align}
\begin{split}
P_{31}^{-}[\hE,\hW,\hF]&=P_{31}^{-}[E,W,F]-\frac{W'_{2,1}}{W_{2,1}}w'(q),\\
P_{32}^{-}[\hE,\hW,\hF]&=P_{32}^{-}[E,W,F]+\left(\frac{W'_{2,1}}{W_{2,1}}
 -\frac{\tvph'_{1}}{\tvph_{1}}\right)w'(q),\\
P_{33}^{-}[\hE,\hW,\hF]&=P_{33}^{-}[E,W,F]+\frac{\tvph'_{1}}{\tvph_{1}}w'(q),
\end{split}
\label{eq:P3i}
\end{align}
which can be regarded as a generalization of (5.1) in Ref.~\cite{BT10}
in the case of type A. It indicates in particular that there is an
infinite number of different factorizations of $P_{3}^{-}$.

The $GL(3,\bbC)$ transformations of the functions $A(z)$, $B(z)$, and
$C(z)$ are calculated as follows. Applying the formulas (\ref{eq:f3}) to
the expression (\ref{eq:hABC1}), we have
\begin{align}
\begin{split}
\hA(w)f''(w)&=\bxi^{\rmT}(w)\bLamb^{-1}\bOm\bLamb\bvphi_{0}(w),\\
\hB(w)&=-\bzeta_{0}^{\rmT}(w)\bLamb^{-1}\bOm\bLamb\bvphi_{0}(w),\\
\hC(w)&=\bzeta_{0}^{\prime\rmT}(w)\bLamb^{-1}\bOm\bLamb\bvphi_{0}(w).
\end{split}
\label{eq:hABC2}
\end{align}
Comparing them with (\ref{eq:ABC1}), we obtain the $GL(3,\bbC)$
transformation of the set of parameters $\{a_{i},b_{i},c_{i}\}$ to
$\{\hat{a}_{i},\hat{b}_{i},\hat{c}_{i}\}$ as
\begin{align}
\widehat{\bOm}=\left(
 \begin{array}{ccc}
 \hat{c}_{0} & \hat{c}_{1} & \hat{c}_{2}\\
 \hat{b}_{0} & \hat{b}_{1} & \hat{b}_{2}\\
 \hat{a}_{0} & \hat{a}_{1} & \hat{a}_{2}
 \end{array}\right)=\bLamb^{-1}\bOm\bLamb,
\label{eq:adj}
\end{align}
which means that $\bOm$ transforms as an adjoint representation.

\section{Type B $3$-fold Superalgebra}
\label{sec:B3SA}

Another notable feature in $\cN$-fold SUSY is that the anti-commutator
of $\cN$-fold supercharges is a polynomial of degree $\cN$ in the
corresponding superHamiltonian~\cite{AST01b,AS03} which constitutes
a generalized superalgebra, called \emph{$\cN$-fold superalgebra}.
The emerged $\cN$th-degree polynomial determines the spectrum of
$H^{\pm}$ in the respective solvable sectors $\cV_{\cN}^{\pm}$ by its
$\cN$ roots. In addition, given that there is a symmetry in an
$\cN$-fold SUSY system, each coefficient of the polynomial is composed
of an invariant quantity of the symmetry. For instance, type A $\cN$-fold
SUSY for any $\cN\in\bbN$ has the $GL(2,\bbC)$ symmetry composed of
linear fractional transformations, and as a consequence every
coefficients of the polynomial involved in type A $\cN$-fold
superalgebra are composed of algebraic invariants of the $GL(2,\bbC)$
transformations~\cite{Ta03a}, cf.\ Section~\ref{sec:Alim}.

In our present type B $3$-fold case, the third-degree polynomial
to be appeared is calculated directly via e.g., $P_{3}^{-}P_{3}^{+}$,
by using (\ref{eq:P3-}). In practice, it is easier to carry out the
calculation in the gauged $z$-space by noting the fact that any
algebraic relation among operators is preserved by a gauge (similarity)
transformation. A direct calculation shows
\begin{align}
\tP_{3}^{+}\tP_{3}^{-}&=-(z')^{3}\left(\frac{\rmd}{\rmd z}+\frac{2A'+B}{A}
 \right)^{2}\left(\frac{\rmd}{\rmd z}+\frac{2A'+B}{A}
 +\frac{f'''}{f''}\right)\left(\frac{\rmd}{\rmd z}-\frac{f'''}{f''}
 \right)\frac{\rmd^{2}}{\rmd z^{2}}\notag\\
&=8\left[\bigl(\tH^{-}+C_{0}(\bOm)\bigr)^{3}+C_{1}(\bOm)\bigl(
 \tH^{-}+C_{0}(\bOm)\bigr)+C_{2}(\bOm)\right],
\label{eq:B3Al}
\end{align}
where the constants $C_{i}(\bOm)$ ($i=0,1,2$) are expressed in terms of
the set of parameters $\{a_{i},b_{i},c_{i}\}$ as
\begin{align}
3C_{0}(\bOm)=&\;a_{2}+b_{1}+c_{0},
\label{eq:C0}\\
3C_{1}(\bOm)=&-(a_{2})^{2}+a_{2}b_{1}-3a_{1}b_{2}+a_{2}c_{0}
 -3a_{0}c_{2}-(b_{1})^{2}+b_{1}c_{0}\notag\\
&-3b_{0}c_{1}-(c_{0})^{2},\\
27C_{2}(\bOm)=&\;2(a_{2})^{3}-3(a_{2})^{2}b_{1}-3(a_{2})^{2}c_{0}
 +9a_{1}a_{2}b_{2}+9a_{0}a_{2}c_{2}-3a_{2}(b_{1})^{2}\notag\\
&+12a_{2}b_{1}c_{0}-18a_{2}b_{0}c_{1}-3a_{2}
 (c_{0})^{2}-18a_{1}b_{2}c_{0}+9a_{1}b_{1}b_{2}\notag\\
&+27a_{1}b_{0}c_{2}+27a_{0}b_{2}c_{1}-18a_{0}b_{1}c_{2}+9a_{0}c_{0}c_{2}
 +2(b_{1})^{3}\notag\\
&-3(b_{1})^{2}c_{0}+9b_{0}b_{1}c_{1}-3b_{1}(c_{0})^{2}+9b_{0}c_{0}c_{1}
 +2(c_{0})^{3}.
\label{eq:C2}
\end{align}

Next, we shall consider the effect of the $GL(3,\bbC)$ transformation on
the operator identity (\ref{eq:B3Al}). Given that $\tP_{3}^{\pm}$ and
$\tH^{-}$ are all transformed according to the rule (\ref{eq:opt}), it is
evident that the algebraic relation among them is maintained under the
transformation and thus the same relation among the transformed quantities
$\htP{}_{3}^{\pm}$ and $\htH{}^{-}$ holds. On the other hand, the parameter
set $\{a_{i},b_{i},c_{i}\}$ also transforms according to (\ref{eq:adj}).
Hence, we must have $C_{i}(\hbOm)=C_{i}(\bOm)$ ($i=0,1,2$) so that
\begin{align}
\htP{}_{3}^{+}\htP{}_{3}^{-}=8\left[\bigl(\htH{}^{-}+C_{0}(\hbOm)\bigr)^{3}
 +C_{1}(\hbOm)\bigl(\htH{}^{-}+C_{0}(\hbOm)\bigr)+C_{2}(\hbOm)\right],
\end{align}
holds. Actually, the latter fact can be immediately derived if we notice
that the constants $C_{i}(\bOm)$ ($i=0,1,2$) are all expressible in terms
of invariants of a matrix under a similarity transformation, namely,
traces and determinants, as
\begin{align}
\begin{split}
&3C_{0}(\bOm)=\Tr\bOm,\qquad
 3C_{1}(\bOm)=-(\Tr\bOm)^{2}+3(\det\bOm)\Tr\bOm^{-1},\\
&27C_{2}(\bOm)=2(\Tr\bOm)^{3}-9(\det\bOm)(\Tr\bOm)\Tr\bOm^{-1}
 +27\det\bOm.
\end{split}
\label{eq:Binv}
\end{align}
This result generalizes the fact in the type A case that every coefficient
of the polynomials appeared in an $\cN$-fold superalgebra consists of
solely algebraic invariants.

\section{Type A Limit and the Subgroup $GL(2,\bbC)$}
\label{sec:Alim}

In the case of type A $\cN$-fold SUSY, it is invariant under the
$GL(2,\bbC)$ linear fractional transformations for arbitrary
$\cN\in\bbN$~\cite{Ta03a}. On the other hand, type A $3$-fold SUSY is
a particular case of type B one, and hence it is evident that the former
must admit the larger $GL(3,\bbC)$ symmetry discussed in the preceding
sections. In this section, we shall show that the $GL(2,\bbC)$ symmetry
in type A $3$-fold SUSY is actually a subgroup of the $GL(3,\bbC)$
and thus any result followed from the former can be reproduced by the
present latter formulation.

In the gauged $z$-space, any type A $\cN$-fold SUSY system is
characterized by two polynomials $A^{(\rmA)}(z)$ and $Q^{(\rmA)}(z)$,
one is of fourth degree and another of second:
\begin{align}
&A^{(\rmA)}(z)=a_{4}^{(\rmA)}z^{4}+a_{3}^{(\rmA)}z^{3}
+a_{2}^{(\rmA)}z^{2}+a_{1}^{(\rmA)}z+a_{0}^{(\rmA)},\\
&Q^{(\rmA)}(z)=b_{2}^{(\rmA)}z^{2}+b_{1}^{(\rmA)}z+b_{0}^{(\rmA)}.
\end{align}
For $\cN=3$, the corresponding gauged Hamiltonian $\tH^{-}$ is expressed
in terms of them as
\begin{align}
\tH^{-}[z]=&-A^{(\rmA)}(z)\frac{\rmd^{2}}{\rmd z^{2}}-\left(Q^{(\rmA)}(z)
 -\frac{A^{(\rmA)\prime}(z)}{2}\right)\frac{\rmd}{\rmd z}\notag\\
&-\frac{A^{(\rmA)\prime\prime}(z)}{6}+Q^{(\rmA)\prime}(z)-R^{(\rmA)},
\label{eq:tH-A}
\end{align}
where $R^{(\rmA)}$ is a constant. Type A $3$-fold SUSY is a special case
of type B one realized by setting $f(z)=z^{2}$~\cite{Ta13}. By comparison
between (\ref{eq:tH-B}) with (\ref{eq:Az})--(\ref{eq:Cz}) where $f(z)=z^{2}$
is substituted and (\ref{eq:tH-A}), each of the coefficients $a_{i}^{(\rmA)}$
($i=0,\dots,4$), $b_{i}^{(\rmA)}$ ($i=0,1,2$), and $R^{(\rmA)}$ are related
to the parameters in type B as
\begin{align}
\begin{split}
&a_{2}=2b_{1}^{(\rmA)}+c_{0},\quad a_{1}=a_{1}^{(\rmA)}+2b_{0}^{(\rmA)},\quad
 a_{0}=2a_{0}^{(\rmA)},\\
&2b_{2}=-a_{3}^{(\rmA)}-2b_{2}^{(\rmA)},\quad b_{1}=-a_{2}^{(\rmA)}
 +b_{1}^{(\rmA)}+c_{0},\quad 2b_{0}=-a_{1}^{(\rmA)}+2b_{0}^{(\rmA)},\\
&c_{2}=2a_{4}^{(\rmA)},\quad c_{1}=a_{3}^{(\rmA)}-2b_{2}^{(\rmA)},\quad
 3c_{0}=a_{2}^{(\rmA)}-3b_{1}^{(\rmA)}+3R^{(\rmA)}.
\end{split}
\label{eq:relAB}
\end{align}
Substituting (\ref{eq:relAB}) into (\ref{eq:C0})--(\ref{eq:C2}), we
see that the three constants $C_{i}(\bOm)$ ($i=0,1,2$) which characterize
3-fold superalgebra of type B reduce to the correct type A formulas
in~\cite{Ta03a}:
\begin{align}
\begin{split}
&C_{0}(\bOm)=R^{(\rmA)},\qquad 3C_{1}(\bOm)=-i_{2}[A^{(\rmA)}]
 +3D_{2}[Q^{(\rmA)}],\\
&27C_{2}(\bOm)=2j_{3}[A^{(\rmA)}]+18I_{1,2}[A^{(\rmA)},Q^{(\rmA)}],
\end{split}
\end{align}
where $D_{2}[Q]$, $i_{2}[A]$, $j_{3}[A]$, and $I_{1,2}[A,Q]$ are the
algebraic invariants, called \emph{transvectants}, composed of fourth-
and second-degree polynomials $A$ and $Q$ (see, e.g., \cite{Gu64} and
references in~\cite{GKO93} for details), and are given in terms of their
respective coefficients $a_{i}$ ($i=0,\dots,4$) and $b_{i}$ ($i=0,1,2$) by
\begin{align}
\begin{split}
&D_{2}[Q]=4b_{0}b_{2}-(b_{1})^{2},\qquad
 i_{2}[A]=12a_{0}a_{4}-3a_{1}a_{3}+(a_{2})^{2},\\
&2j_{3}[A]=72a_{0}a_{2}a_{4}-27a_{0}(a_{3})^{2}-27(a_{1})^{2}a_{4}
 +9a_{1}a_{2}a_{3}-2(a_{2})^{3},\\
&I_{1,2}[A,Q]=6a_{4}(b_{0})^{2}-3a_{3}b_{0}b_{1}+2a_{2}b_{0}b_{2}
 +a_{2}(b_{1})^{2}-3a_{1}b_{1}b_{2}+6a_{0}(b_{0})^{2},
\end{split}
\end{align}
where the superscript (A) has been omitted for the simplicity.\\

In Ref.~\cite{BT10}, we considered the $GL(2,\bbC)$ linear fractional 
transformation
\begin{align}
z=\frac{\alpha w+\beta}{\gamma w+\delta},\qquad(\alpha,\beta,\gamma,\delta
 \in\bbC,\quad\Delta=\alpha\delta-\beta\gamma\neq 0).
\end{align}
The three-dimensional type A monomial subspace is transformed as
\begin{align}
\tcV_{3}^{(\rmA)}[z]=\braket{1,z,z^{2}}\ \to\ \htcV{}_{3}^{-}[w]&=
 (\gamma w+\delta)^{2}\tcV_{3}^{(\rmA)}[z]\Bigr|_{z=(\alpha w+\beta)/
 (\gamma w+\delta)}\notag\\
&=\braket{(\gamma w+\delta)^{2},(\alpha w+\beta)(\gamma w+\delta),
 (\alpha w+\beta)^{2}}\notag\\
&=\tcV_{3}^{(\rmA)}[w].
\end{align}
Comparing it with (\ref{eq:trB}) and (\ref{eq:GL3}), and noting
$f(w)=w^{2}$ in the type A case, we recognize that the above $GL(2,\bbC)$
transformation is in fact a subgroup of the $GL(3,\bbC)$ transformation
(\ref{eq:GL3}) characterized by the following specific form of $\bLamb$:
\begin{align}
\bLamb=\left(
 \begin{array}{ccc}
 \delta^{2} & 2\gamma\delta & \gamma^{2}\\
 \beta\delta & \alpha\delta+\beta\gamma & \alpha\gamma\\
 \beta^{2} & 2\alpha\beta & \alpha^{2}
 \end{array}
 \right).
\end{align}
The Wronskian $W_{2,1}(w)$ in this case is calculated as
\begin{align}
W_{2,1}(w)=\Delta(\gamma^{2}w^{2}+2\gamma\delta w+\delta^{2})
 =\Delta\tvph_{1}(w).
\end{align}
Hence, the $GL(2,\bbC)$ transformations of $E(q)$, $W(q)$, and $F(q)$
in (\ref{eq:t1}), (\ref{eq:t2}), and (\ref{eq:t3}), respectively,
as a subgroup of the $GL(3,\bbC)$ one, namely, read as
\begin{align}
\begin{split}
&\hE(q)=E(q)+\frac{2\gamma w'(q)}{\gamma w(q)+\delta}=E(q)
 -\frac{2\gamma z'(q)}{\gamma z(q)-\alpha},\\
&\hW(q)=W(q),\qquad \hF(q)=F(q)=0,
\end{split}
\end{align}
which coincides exactly with (2.12) in~\cite{BT10}.
In particular, each factor of the factorized type A 3-fold supercharge
$P_{3}^{-}=P_{31}^{-}P_{32}^{-}P_{33}^{-}$ is transformed as
\begin{align}
\begin{split}
P_{31}^{-}[\hW,\hE,0]&=P_{31}^{-}[W,E,0]+\frac{2\gamma z'(q)}{
 \gamma z(q)-\alpha},\\
P_{32}^{-}[\hW,\hE,0]&=P_{32}^{-}[W,E,0],\\
P_{33}^{-}[\hW,\hE,0]&=P_{33}^{-}[W,E,0]-\frac{2\gamma z'(q)}{
 \gamma z(q)-\alpha},
\end{split}
\end{align}
and thus (5.1) of~\cite{BT10} is reproduced.

As a final remark, we note that the three invariants in (\ref{eq:inv})
reduce to the two ones of the $GL(2,\bbC)$ in the case of type A 3-fold
SUSY (cf., Eqs.~(2,12) and (2.13) in Ref.~\cite{BT10}) in the type A
limit $F\to 0$:
\begin{align}
I_{1}[E,W,0]=W,\quad I_{2}[E,W,0]=2E'-E^{2},\quad I_{3}[E,W,0]=0.
\end{align}
The $GL(2,\bbC)$ considered in the type A case is a subgroup of the
present $GL(3,\bbC)$, and thus a consistency has been checked.

\section{$GL(\cN,\bbC)$ Invariance of $\cN$-fold SUSY}
\label{sec:GLNC}

By generalizing the argument about the $GL(3,\bbC)$ invariance of
type B 3-fold SUSY in Section~\ref{sec:GL3C}, we easily come to the
conclusion that any $\cN$-fold SUSY system has $GL(\cN,\bbC)$ invariance.
As shown in \cite{GT05}, a specific $\cN$-dimensional linear function space
\begin{align}
\tcV_{\cN}^{-}[z]=\braket{\tvph_{1}(z),\dots,\tvph_{\cN}(z)},
\end{align}
uniquely determines an $\cN$-fold SUSY system and vice versa.
On the other hand, under any $GL(\cN,\bbC)$ transformation on
$\tcV_{\cN}^{-}[z]$ defined by
\begin{align}
\htvph_{i}(w)=\sum_{j=1}^{\cN}\lambda_{ij}\tvph_{j}(w),\quad(i=1,\dots,\cN),
\end{align}
it is evident that the vector space $\tcV_{\cN}^{-}[z]$ is invariant:
\begin{align}
\htcV{}_{\cN}^{-}[w]=\braket{\htvph_{1}(w),\dots,\htvph_{\cN}(w)}
 =\tcV_{\cN}^{-}[w].
\end{align}
Hence, the assertion must follow in the same sense as in
Section~\ref{sec:GL3C}.

The argument about the subgroup $GL(2,\bbC)$ in the limiting case of
type A in Section~\ref{sec:Alim} further suggests that the $GL(2,\bbC)$
linear fractional transformation on any type A $\cN$-fold SUSY system
for an arbitrary $\cN\in\bbN$ considered in \cite{Ta03a} is also
a subgroup of the above symmetry group $GL(\cN,\bbC)$. In fact, the
$GL(2,\bbC)$ transformation acts on the type A solvable sector
$\tcV_{\cN}^{(\rmA)}[z]=\braket{1,z,\dots,z^{\cN-1}}$ as
\begin{align}
\tcV_{\cN}^{(\rmA)}[z]\ \to\ \htcV{}_{\cN}^{-}[w]&=(\gamma w+\delta)^{\cN-1}
 \tcV_{3}^{(\rmA)}[z]\Bigr|_{z=(\alpha w+\beta)/(\gamma w+\delta)}\notag\\
&=\braket{(\gamma w+\delta)^{\cN-1},(\alpha w+\beta)(\gamma w+\delta)^{\cN-2},
 \cdots,(\alpha w+\beta)^{\cN-1}}\notag\\
&=\tcV_{3}^{(\rmA)}[w],
\end{align}
and thus it is a particular $GL(\cN,\bbC)$ transformation:
\begin{align}
\htvph_{i}(w)=\sum_{j=1}^{\cN}\lambda_{ij}w^{j-1},\quad(i=1,\dots,\cN),
\end{align}
where the coefficient $\lambda_{ij}$ ($i,j=1,\dots,\cN$) is explicitly
given by
\begin{align}
\lambda_{ij}=\sum_{k=0}^{j-1}\left(\begin{array}{c}\cN-i\\ j-k-1\end{array}
 \right)\left(\begin{array}{c}i-1\\ k\end{array}\right)\alpha^{k}
 \beta^{i-k-1}\gamma^{j-k-1}\delta^{\cN-i-j+k+1}.
\end{align}
In the above, any binomial coefficient $\left(\begin{array}{c}m\\ 
n\end{array}\right)$ is understood to be zero for $m<n$.

\section{Discussion and Summary}
\label{sec:discus}

In this paper, we have developed the effects of a general variable
transformation in type B 3-fold SUSY and then applied them to
the $GL(3,\bbC)$ homogeneous linear transformation. Then, we have
shown that there are three invariant functions $I_{i}(q)$ ($i=1,2,3$)
in (\ref{eq:inv}) composed of the three functions $E(q)$, $W(q)$, and
$F(q)$ which characterize a type B 3-fold SUSY system, and that such
a system is completely expressible in terms of them, (\ref{eq:invP})
and (\ref{eq:invV}), and hence is invariant under the $GL(3,\bbC)$.
We have also shown that the parameter set $\{a_{i},b_{i},c_{i}\}$
transform as an adjoint representation (\ref{eq:adj}).
We have calculated the characteristic polynomial of third degree
emerged from the product of type B 3-fold supercharges and then
confirmed that its coefficients are all $GL(3,\bbC)$ invariants
(\ref{eq:Binv}).
We have considered the type A limit and shown that the $GL(2,\bbC)$
linear fractional transformation which leaves any type A system
invariant is a subgroup of the $GL(3,\bbC)$. In the last, we have
argued that any $\cN$-fold SUSY system has invariance under
a $GL(\cN,\bbC)$ transformation for any $\cN\in\bbN$.

The present results have several applications in the future development.
Noting first the fact that the $GL(2,\bbC)$ invariance was so efficient
in classifying the type A $\cN$-fold SUSY models in~\cite{Ta03a}, we
expect that the present $GL(3,\bbC)$ will enable one to investigate
systematically what kind of type B 3-fold SUSY potentials can be
realized. In this respect, it would be also interesting to reconsider
the classification of type A 3-fold SUSY in view of the more general
$GL(3,\bbC)$ than the previous $GL(2,\bbC)$. The latter study would
further give a suggestion on what can be anticipated if we review type A
$\cN$-fold SUSY for any $\cN\in\bbN$ by utilizing a $GL(cN,\bbC)$
transformation.

Another important aspect of the transformation properties is that
each of factorized components of the type B 3-fold supercharge is
\emph{not} invariant under the $GL(3,\bbC)$, cf.\ (\ref{eq:P3i}).
In the type A 2- and 3-fold SUSY cases, it was shown in~\cite{BT09,BT10}
that the non-invariance resulted in the existence of different sets
of intermediate Hamiltonians. In the study of the number and 
classification of such inequivalent sets, the $GL(2,\bbC)$ transformation
played a central and crucial role. Hence, we expect that the present
$GL(3,\bbC)$ transformation will also provide an indispensable tool for
investigating both the existence and classification of different sets
of intermediate Hamiltonians in the most general type B 3-fold SUSY
system.

Regarding the subject of intermediate Hamiltonians in $\cN$-fold SUSY,
we would like to recall its relevance on the concept of \emph{shape
invariance}~\cite{Ge83} and its \emph{muti-step} generalization~\cite{
BDGKPS93}, which have been practical methods in constructing solvable
quantum Hamiltonians. Not only was ordinary shape invariance treated
efficiently in the type A 2- and 3-fold SUSY with intermediate
Hamiltonians~\cite{BT09,BT10}, but two-step one was also classified
systematically in the framework of type A 2-fold SUSY~\cite{RT13}.
The next scope in this field is definitely three-step generalization,
and our prospect is that the type B 3-fold SUSY formalism with
the present $GL(3,\bbC)$ equipment would allow us to develop
farsightedly the issue.

\begin{acknowledgments}
This work was motivated during a collaborative work with B.~Bagchi.
We would like to express our gratitude to him for the fruitful discussions.

\end{acknowledgments}

\appendix

\section{Formulas for the Transformation}
\label{app:forms}

The general transformation from the $z$-space to the $w$-space in
Section~\ref{sec:GTB3} is carried out based on the formulas:
\begin{align}
z=\tvph_{2}(w)/\tvph_{1}(w),\qquad f(z)=\tvph_{3}(w)/\tvph_{1}(w).
\label{eq:f0}
\end{align}
Using them, we can calculate derivatives of $z$ and $f(z)$ with
respect to $w$ as
\begin{align}
\begin{split}
&\frac{\rmd z}{\rmd w}=\frac{W_{2,1}(w)}{\tvph_{1}(w)^{2}},\qquad
 \frac{\rmd^{2}z}{\rmd w^{2}}=\frac{W'_{2,1}(w)}{\tvph_{1}(w)^{2}}
 -\frac{2W_{2,1}(w)\tvph'_{1}(w)}{\tvph_{1}(w)^{3}},\\
&\frac{\rmd^{3}z}{\rmd w^{3}}=\frac{W''_{2,1}(w)}{\tvph_{1}(w)^{2}}
 -\frac{4W'_{2,1}(w)\tvph'_{1}(w)+2W_{2,1}(w)\tvph''_{1}(w)}{\tvph_{1}(w)^{3}}
 +\frac{6W_{2,1}(w)\tvph'_{1}(w)^{2}}{\tvph_{1}(w)^{4}},\\
&\frac{\rmd f(z)}{\rmd w}=\frac{W_{3,1}(w)}{\tvph_{1}(w)^{2}},\qquad
 \frac{\rmd^{2}f(z)}{\rmd w^{2}}=\frac{W'_{3,1}(w)}{\tvph_{1}(w)^{2}}
 -\frac{2W_{3,1}(w)\tvph'_{1}(w)}{\tvph_{1}(w)^{3}},\\
&\frac{\rmd^{3}f(z)}{\rmd w^{3}}=\frac{W''_{3,1}(q)}{\tvph_{1}(q)^{2}}
 -\frac{4W'_{3,1}(w)\tvph'_{1}(w)+2W_{3,1}(w)\tvph''_{1}(w)}{\tvph_{1}(w)^{3}}
 +\frac{6W_{3,1}(w)\tvph'_{1}(w)^{2}}{\tvph_{1}(w)^{4}},
\end{split}
\label{eq:f1}
\end{align}
where and throughout the paper we have employed the following
notation for the Wronskians:
\begin{align}
&W_{i,j}(w)=\tvph'_{i}(w)\tvph_{j}(w)-\tvph_{i}(w)\tvph'_{j}(w),
\label{eq:Wron1}\\
&W_{i',j'}(w)=\tvph''_{i}(w)\tvph'_{j}(w)-\tvph'_{i}(w)\tvph''_{j}(w),
\label{eq:Wron2}\\
&W_{ij,kl}(w)=W'_{i,j}(w)W_{k,l}(w)-W_{i,j}(w)W'_{k,l}(w).
\label{eq:Wron3}
\end{align}
With these formulas, we obtain
\begin{align}
\begin{split}
z'(q)=&\;\frac{W_{2,1}(w)}{\tvph_{1}(w)^{2}}u'(q),\qquad
 zf'(z)-f(z)=\frac{W_{3,2}(w)}{W_{2,1}(w)},\\
z''(q)=&\;\frac{W_{2,1}(w)}{\tvph_{1}(w)^{2}}u''(q)
 +\left(\frac{W'_{2,1}(w)}{\tvph_{1}(w)^{2}}
 -\frac{2W_{2,1}(w)\tvph'_{1}(w)}{\tvph_{1}(w)^{3}}\right)u'(q)^{2},\\
f'(z)=&\;\frac{W_{3,1}(w)}{W_{2,1}(w)},\qquad
 f''(z)=\frac{W_{31,21}(w)\tvph_{1}(w)^{2}}{W_{2,1}(w)^{3}},\\
f'''(z)=&\;\frac{W'_{31,21}(w)\tvph_{1}(w)+2W_{31,21}(w)\tvph'_{1}(w)}{
 W_{2,1}(w)^{4}}\tvph_{1}(w)^{3}\\
&-\frac{3W_{31,21}(w)W'_{2,1}(w)\tvph_{1}(w)^{4}}{W_{2,1}(w)^{5}}.
\end{split}
\label{eq:f2}
\end{align}

For the $GL(3,\bbC)$ transformation in Section~\ref{sec:GL3C},
we need to calculate several Wronskians defined in
(\ref{eq:Wron1})--(\ref{eq:Wron3}). Each component of the
transformation (\ref{eq:GL3}) is
\begin{align}
\tvph_{i}(w)=\lambda_{i1}+\lambda_{i2}w+\lambda_{i3}f(w).
\end{align}
Then, we have the following:
\begin{align}
\begin{split}
&W_{2,1}(w)=\blambda_{33}-\blambda_{32}f'(w)+\blambda_{31}(wf'(w)-f(w)),\\
&W_{3,1}(w)=-\blambda_{23}+\blambda_{22}f'(w)-\blambda_{21}(wf'(w)-f(w)),\\
&W_{3,2}(w)=\blambda_{13}-\blambda_{12}f'(w)+\blambda_{11}(wf'(w)-f(w)),\\
&W_{31,21}(w)=(\det\bLamb)\tvph_{1}(w)f''(w),\qquad
 W_{2',1'}(w)=\blambda_{31}f''(w),\\
&W_{3',1'}(w)=-\blambda_{21}f''(w),\qquad W_{3',2'}(w)=\blambda_{11}f''(w),
\end{split}
\label{eq:f3}
\end{align}
where $\blambda_{ij}$ is the cofactor of the matrix element $\lambda_{ij}$
in $\bOm$.


\bibliography{refsels}
\bibliographystyle{npb}



\end{document}